\documentclass[12pt]{article}
\usepackage{amsmath,amssymb,bm}

\newcommand{\CC}{\mathbb{C}}
\newcommand{\RR}{\mathbb{R}}
\newcommand{\ZZ}{\mathbb{Z}}
\newcommand{\Hil}{\mathcal{H}}
\newcommand{\ket}[1]{| #1 \rangle}
\newcommand{\bra}[1]{\langle #1 |}
\newcommand{\Eq}[1]{Eq.~\ref{#1}}
\newcommand{\Eqs}[1]{Eqs.~{#1}}
\newcommand{\Sec}[1]{Sect.~\ref{Sec:#1}}
\newcommand{\Op}[1]{\hat{#1}}
\newcommand{\Id}{\mathbf 1}
\newcommand{\OId}{\Op{I}}
\newcommand{\e}{\mathfrak{e}}
\newcommand{\X}[1]{\Op{\sigma}^{x\vphantom{y}}_{#1}}
\newcommand{\Y}[1]{\Op{\sigma}^{y}_{#1}}
\newcommand{\Z}[1]{\Op{\sigma}^{z\vphantom{y}}_{#1}}
\newcommand{\mi}{{\rm i}}
\newcommand{\me}{{\rm e}}
\newcommand{\Cl}{C\!\ell}
\newcommand{\spn}[1]{\mathcal{#1}}
\newcommand{\chn}[1]{\underline{#1}}

\newcommand{\mt}[1]{\bm{#1}}
\newcommand{\vac}{\emptyset}
\DeclareMathOperator{\Tr}{Tr}
\DeclareMathOperator{\Ad}{Ad}

\begin{document}

\title{Effective Simulation of State Distribution in Qubit Chains}

\author{Alexander Yu.\ Vlasov\footnote{qubeat@mail.ru}\\
\small Federal Radiology Center (IRH)\\
\small 197101, Mira Street 8, Saint Petersburg, Russia}

\date{}

\maketitle

\sloppy

\begin{abstract}
This work recollects a non-universal set of quantum gates described by higher-dimensional 
Spin groups. They are also directly related with matchgates in theory of quantum computations 
and complexity. Various processes of quantum state distribution along a chain such as perfect 
state transfer and different types of quantum walks can be effectively modeled on classical 
computer using such approach.
\end{abstract}

\section{Introduction}
\label{Sec:intro}

The non-universal sets of quantum gates discussed in this work are known 
in the wide variety of contexts from matchgates in theory of complexity to 
Majorana modes in the solid body physics 
\cite{Val1,TD2,Kni1,Kit00,BK00,Vla0,Joz8,Joz9,JMS15,Brod16}, but they 
were often considered rather as {\em possible obstacles} for construction of 
general purpose quantum computers. 

For example, such a non-universal set was described by the author as some auxiliary 
effect of construction of the universal set of quantum gates \cite{Vla0} inspired
by an earlier work with an application of spinors and Clifford algebra formalism \cite{Vla99}.
In the works about Majorana modes \cite{Kit00,BK00} similar non-universal quantum gates 
are described by Hamiltonians with second-order terms and it was 
noted about ``physical implementation'' that an additional ``four-particle interaction'' 
necessary for universal quantum computation ``will be particularly difficult to realize.''

However, the model under consideration can be treated in the {\em more constructive way}.
The matchgates was initially introduced in the theory of complexity as a special set 
of quantum gates simulated classically in polynomial time \cite{Val1}
and equivalence of the nearest-neighbor (n.n.) unitary matchgates with the non-universal 
set discussed above was proved soon \cite{TD2,Kni1}. 

Here the term ``matchgates'' is often used for historical reasons and 
for the consistency with works of other authors. The spinor representation 
\cite{Vla0} reintroduced in \Sec{descr} can be more 
convenient for the purposes of presented work. 
A similar approach with $2^n$-components spinors was also briefly mentioned in \cite{Wilc09}
in relation with Majorana modes and `topological quantum computations' \cite{Kit00,top08}.
 
The spinoral approach also let us avoid necessity to digress into explanation
of some special cases irrelevant to presented work.
For example, without n.n. condition the matchgates can perform 
universal quantum computation \cite{TD2,Joz8} that may not be effectively
simulated classically. Originally suggested in \cite{Val1} nearest-neighbor 
matchgates also could include non-physical (non-unitary and non-invertible) 
gates that should be discussed elsewhere \cite{JMS15}. 
 
The considered models are relevant not only to effective {\em classical}
simulations of matchgate circuits.
An equivalence of match-circuits of width $n$ and an universal {\em quantum} 
computer with ``exponentially compressed'' number of qubits $O(log(n))$ was shown in \cite{Joz9}. 

Thus, many problems in the theory of quantum information let us also consider 
the possibility of effective implementation of some restricted set of quantum 
circuits with such gates {\em as some benefit} for simulation, 
comparison of classical and quantum computational complexity 
and other tasks. 

An application of this particular class of effectively
simulated quantum circuits as a model of state distribution 
along a chain of qubits is considered in this work.
The different types of quantum chains are introduced in \Sec{qchain} together with examples
of processes appropriate for effective modeling such as perfect state transfer 
and different types of continuous and discrete quantum walks. 

The {\em local representation} of a single link is discussed in \Sec{cmp} 
for establishing relation between a qubit chain and a simpler {\em scalar} model with 
dimension of Hilbert space is equal to number of nodes.
It is used further in \Sec{descr} for a {\em spinoral} 
description of Hamiltonians and the quantum evolution of the entire chain. 
The applications to the state distribution are summarized 
and extended for the multi-particle case in \Sec{effmod}.

\section{Quantum chains}
\label{Sec:qchain}

\subsection{Chain types}
\label{Sec:chtyp}

A {\em chain} is defined by a set of nodes $k = 1,\dots,n$ together with 
the pairs for representation of links.
The usual (linear) chain is defined by the links $(k,k+1)$, $k=1,\dots,n-1$. 
A {\em ring}, {\em i.e.}, {\em circular chain} has an additional link $(n,1)$.

The {\em qubit chain} is formally equivalent to a quantum computational network with 
$n$ qubits and quantum gates acting either on the single qubit or on the nearest neighbors
$(k,k+1)$, $k=1,\dots,n-1$. 
Additional two-gates for the pair of qubits $(1,n)$ at the ends of the chain
produces the {\em qubit ring}. 
For the systems with $n$ qubits the dimension of a space of states
$\Hil = \Hil_2 \otimes \cdots \otimes \Hil_2$ is $2^n$ and methods of effective 
classical simulation should not work {\em directly} with this exponentially large space.

A couple of related models is also considered together with the chain of qubits in this work.
The {\em scalar quantum chains and rings} with $n$ nodes are described by $n$-dimensional 
space $\Hil_n$. {\em The coined (discrete) quantum walk} \cite{ADZ93,AB01,Kem03,Ken06,V-A12} 
is equipped with two-dimensional {\em control space} and the 
space of states for this model is $\Hil_2 \otimes \Hil_n  \simeq \Hil_{2n}$.

\subsection{Coined quantum walk on scalar chain}
\label{Sec:coin}

The simple models of coined quantum walks on a line or a circle have the straightforward 
implementation using the scalar quantum chain with the doubled number of nodes.
Let us start with the infinite quantum chain $\ket{k}$ with arbitrary integer indexes
of nodes $k \in \ZZ$ to postpone the consideration of boundary conditions.
Two operators for exchanging pairs of nodes may be introduced
\begin{equation}
\Op S_1 \colon \ket{2k} \leftrightarrow \ket{2k+1},\quad
\Op S_2 \colon \ket{2k+1} \leftrightarrow \ket{2k+2}.
\label{S1S2}
\end{equation}
Let us consider the composition $\Op S_{12}$ of the operators $\Op S_1$ and $\Op S_2$
\begin{equation}
\Op S_{12} \colon \begin{cases}
\ket{2k} \rightarrow \ket{2k+1} \rightarrow \ket{2k+2}  \\
 \ket{2k+1} \rightarrow \ket{2k} \rightarrow \ket{2k-1}
 \end{cases}
\label{S12}
\end{equation}
The action of \Eq{S12} can be considered as two independent
shifts with the double step in the opposite directions for odd and even indexes.
For the comparison with the model of a coined quantum walk, the nodes can be
reordered into two infinite chains using the notation 
$\ket{c}\ket{k}$ for $\ket{2k+c}$ where $k \in \ZZ$, $c=0,1$.
In such a case the operation $\Op S_{12}$ \Eq{S12} corresponds to
the transformation
\begin{equation}
\ket{0,k} \rightarrow \ket{0,k+1}, 
\quad \ket{1,k} \rightarrow \ket{1,k-1}.
\label{S12_2}
\end{equation}
and may be rewritten as
\begin{equation}
    \Op S_{12} = \ket{0}\bra{0} \otimes \Op R + \ket{1}\bra{1} \otimes \Op L,
    \label{CondS12}
\end{equation}
where $\Op R$ and $\Op L$ denote right and left shifts $\ket{k} \to \ket{k \pm 1}$
respectively.
 
The operators such as \Eq{CondS12} is an example of {\em conditional quantum dynamics} 
\cite{cond95} with a {\em control qubit} $\ket{c}$ and a chain as a {\em target subsystem}.
The same model is widely used for the {\em coined quantum walk} \cite{ADZ93,AB01,Kem03} 
with {\em coin flips} operators $\Op C$ acting on the control qubit \cite{AB01,Kem03}
\begin{equation}
\Op W = \Op S_{12}(\Op C \otimes \OId),
\label{QW}
\end{equation}
where $\OId$ is the identity (unit) operator on the chain. 
In the initial notation with the single chain (before decomposition on control and target
subsystems) the coin operator $\Op C$ acts on the same pairs of nodes as
$\Op S_1$. So the coined quantum walk may be described by an analogue 
of the two-step method \Eq{S1S2} with $\Op S'_1 = \Op S_1\Op C$ 
also known due to staggered and Szegedy quantum walks \cite{Szeg,stag1,stag,wong}.

\smallskip

A case of a circular quantum chain with $2n$ nodes is analogous with the infinite 
model considered above. The swap transformations $S_1$ and $S_2$ are defined similarly
with \Eq{S1S2} using addition modulo $2n$. The relation between
such a chain and a coined quantum walk on a circle with $n$ nodes again
corresponds to the reordering $\ket{c}\ket{k}$ for $\ket{2k+c}$ where 
$k = 0,\ldots,n-1$, $c=0,1$ and arithmetic modulo $n$ is used for the second index in \Eq{S12_2}.
The expression \Eq{CondS12} should be rewritten
in such a case using $n \times n$ the matrix $\Op U_n$ of the cyclic shift
\begin{equation}
    \Op S_{12} = \ket{0}\bra{0} \otimes \Op U_n + \ket{1}\bra{1} \otimes \Op U^\dag_n.
    \tag{\ref{CondS12}$'$}
    \label{cshift}
\end{equation}

The boundary conditions for a linear chain with $2n$ nodes is a bit more difficult.
Here the operations $\Op S_1$ and $\Op S_2$ in \Eq{S1S2} should be modified
for the ends of the chain. The bounded analogues should not affect nodes
outside of the range $1,\ldots,2n-1$. 
After the transition to the coined quantum walk on a line with $n$ nodes such
an approach produces specific reflection effects on the boundaries.

Very similar models were used for graphs with more general structure
for approaches with staggered and Szegedy quantum walks \cite{Szeg,stag1,stag,wong},
but it is not well suited to the consideration of an effective simulation model
discussed further.

\subsection{Continuous quantum walk on chains}
\label{Sec:cont}

Let us consider an alternative model of the {\em continuous quantum walk}
\cite{Ken06,V-A12,FarGut} also applicable to the consideration of 
the {\em perfect state transport} \cite{Bose03,Perf04,Perf05,dual}.
The methods are relevant both for scalar and qubit chains.

The scalar chain is encoded by qubits adapting of the standard approach from the theory 
of spin waves \cite{FLP3,SpWav09,plinko}. Inside of the whole $2^n$-dimensional Hilbert 
space of system with $n$ qubits a state of the chain can be mapped 
into $n$-dimensional subspace spanned by states with a single unit in 
the computational basis 
\begin{equation}\label{qub2scal}
\ket{\chn k} \equiv \ket{{\underbrace{0\ldots 0}_{k-1}\,}1\underbrace{0\ldots 0}_{n-k}},
\quad k=1,\ldots,n.
\end{equation}

In such a model the element $\Op H_{jk}$ of Hamiltonians
is nonzero only for the linked nodes $j$ and $k$.
A simple example is an adjacency matrix of graphs with 
unit elements for the edges.
For linear and circular chains Hamiltonians defined by an adjacency matrix
are quite simple and may be presented respectively as 
$\Op H^{\rm l}$ and $\Op H^{\rm c}$ with the only nonzero elements
\begin{eqnarray*}
 &\Op H^{\rm l}_{k,k-1} =  \Op H^{\rm l}_{k-1,k} = 1, \quad& k = 1,\ldots,n-1, \\
 &\Op H^{\rm c}_{k \bmod n,k-1} =  \Op H^{\rm c}_{k-1,k \bmod n} = 1, \quad& k = 1,\ldots,n.
\end{eqnarray*}

In the model discussed earlier an evolution of a chain on each discrete step
was represented by the unitary operators $\ket{\psi} \to \Op U \ket{\psi}$. 
A similar approach for models with given Hamiltonian may be developed
by choice of the fixed time step $\tau$. 
The evolution of a chain for each step is 
described by the operator 
\begin{equation}
    \Op U(\tau) = \exp(-\mi \Op H \tau) 
\label{expH}
\end{equation}
and in a general case only for $\tau \to 0$ the structure of the links
represented by the Hamiltonian is simply related with the evolution after
the single step due to \mbox{$\Op U(\tau) \approx \OId - \mi \Op H \tau$}.

Evolution due to the Hamiltonian constructed from the adjacency matrix for 
finite $\tau$ corresponds to a state propagation along connected
paths in the graph described by the links. For the qubit chain such a model may
produce too complex gates acting on arbitrary big number of qubits. 

The problem can be partially resolved by using
short chains with two and three qubits as the building blocks \cite{Perf04}.
On the other hand, some Hamiltonians with the specific values
instead of units in the adjacency matrix may produce the transport 
of a localized state for particular values of $\tau$, {\em i.e.},
after some period of time such an evolution is formally equivalent to 
the swap gate applied to the ends of the chain \cite{Perf04,Perf05}.
For a scalar quantum chain with $n$ nodes a neat example is an
evolution described by the Hamiltonian with only nonzero elements
for adjacent nodes 
\begin{equation}
 \Op H_{k,k-1} =  \Op H_{k-1,k} = \frac{1}{2}\sqrt{k(n-k)}, ~ k = 1,\ldots,n-1.
 \label{HSx}
\end{equation}
The model is quite illustrative due to the possibility to map nodes into states
of a quantum particle with the spin \mbox{$s=(n-1)/2$}. A node with an index $k = 0,\ldots,n-1$
would correspond to the spin projection $s_z=(-s/2+k)\hbar$ on $z$ axis of such
a formal quantum system. The {\em natural system of units} with 
$\hbar = 1$ is used further for simplicity.

Such a Hamiltonian may be simply interpreted as a rotation of a particle with
the spin $s=(n-1)/2$ along $x$ axis \cite{FLP3} and so the state {\em spin up} with
the projection \mbox{$s_z=+s/2$} evolves into the state {\em spin down} $s_z=-s/2$.
The formal spin model is convenient for the consideration of the quantum chain
not only because elements of the Hamiltonian correspond to the next-neighbor 
structure, but also due to the quite straightforward expression for
the evolution \Eq{expH}. 

A spin-half particle is the simple example corresponding to
a chain with two nodes or the qubit $\ket{\psi} = \alpha\ket{0}+\beta\ket{1}$.
The change of a state due to rotation corresponds to some unitary $2 \times 2$
matrix $\ket{\psi'} = \Op U\ket{\psi}$. Let us consider $n$-dimensional linear
space of polynomials of order $n-1$ with two variables 
$\alpha$, $\beta$ with the basis
\begin{equation}
 p_k(\alpha,\beta) = \textstyle\sqrt{\frac{(n-1)!}{k!(n-k-1)!}}\alpha^k\beta^{n-k-1} 
 = \sqrt{C^k_{n-1}} \alpha^k\beta^{n-k-1}, 
\label{polym}
\end{equation}
where  $k = 0, \dots, n-1$.
The transformation of such polynomials due to application of an unitary $2 \times 2$
matrix $U$ to the coefficients $(\alpha,\beta)$ of a qubit mentioned 
earlier describes rotation of the formal system with the spin $s$ \cite{FLP3,Weyl}. 
The coefficients $\sqrt{C}{}^k_{n-1}$ in \Eq{polym} ensure
unitarity of the $n \times n$ transformation matrix.

The initial state $\ket{0}$ with $\alpha=1$, $\beta=0$ produces only 
one nonzero polynomial $p_0$. The quantum {\it NOT} gate transforms  
$\ket{0} \to \ket{1}$ with $\alpha=0$, $\beta=1$ and only nonzero
polynomial $p_{n-1}$. If states of the formal spin system are mapped into
a chain with $n=2s+1$ nodes such a transformation corresponds to
the exchange of values between two end nodes $\ket{1}$ and $\ket{n}$.
On the other hand, the gate may be implemented by 
continuous spin rotation around $x$ axis using the Hamiltonian \Eq{HSx}.

The same gate may be implemented using rotation around $y$ axis.
In such a case all nonzero elements of the Hamiltonian are pure imaginary. 
Models of the state transport often utilize real Hamiltonian such as \Eq{HSx}
and a more general case with complex coefficients is adapted for 
the {\em chiral quantum walks} \cite{Biam13,Lloyd16}.
Here is useful to compare the Hamiltonian $\Op H=\Op H^x$ \Eq{HSx} with
$\Op H^y$ and $\Op H^z$ representing rotations around $y$ and $z$ axis. 
The nonzero elements may be expressed by equations
\begin{eqnarray}
&&\Op H^x_{k,k-1} =  \Op H^x_{k-1,k} = 
\mi \Op H^y_{k,k-1} =  -\mi \Op H^y_{k-1,k}  = \frac{1}{2}\sqrt{k(n-k)}
\nonumber\\
&&\Op H^z_{k,k} = k - \frac{1}{2}(n-1).
\label{HSxyz}
\end{eqnarray}
An arbitrary rotation is represented by the linear combination of the three 
Hamiltonians 
$\Op H_\lambda=\lambda_x\Op H^x+\lambda_y\Op H^y+\lambda_z\Op H^z$, but 
the perfect transfer may be implemented only for rotation around an axis
perpendicular to $z$ for $\lambda_z = 0$.

\section{Comparison of scalar and qubit chains}
\label{Sec:cmp}

Let us consider the relation between scalar and qubit chains. The quantum circuit
model is convenient for the initial consideration of the {\em single link with two nodes}.
The space of states for two qubits is four-dimensional and transitions should not
change the number of units in the computational basis. The nontrivial
evolution is only between $\ket{01}$ and $\ket{10}$, but other two states
may only change phases. 

The matrix of the transformation can be represented in general as
\begin{equation}
M = \begin{pmatrix}
\me^{\mi \nu} & 0 & 0 & 0 \\
0 & \mu_{11} & \mu_{12} & 0 \\
0 & \mu_{21} & \mu_{22} & 0 \\
0 & 0 & 0 & \me^{\mi \nu'} 
\end{pmatrix}
\label{tran4x4}
\end{equation}

Let us recall that the matchgate \cite{Val1,TD2,Kni1,Joz8,Joz9} is a quantum two-gate
with $4 \times 4$ matrix composed from elements of two 
matrices $A, B \in \mathrm{SU}(2)$
\begin{equation}
\me^{\mi\theta}M(A,B) = \begin{pmatrix}
A_{11} & 0 & 0 & A_{12} \\
0 & B_{11} & B_{12} & 0 \\
0 & B_{21} & B_{22}  & 0 \\
A_{21} & 0 & 0 & A_{22}. 
\end{pmatrix}
\label{match}
\end{equation}
The matrix of the transformation \Eq{tran4x4} corresponds
to \Eq{match} for \mbox{$\theta = -(\nu+\nu')/2$} and
\begin{subequations}\label{matchtran}
\begin{gather}
A = \begin{pmatrix} \me^{\mi (\nu-\nu')/2} & 0 
    \\ 0 & \me^{\mi (\nu'-\nu)/2}
    \end{pmatrix}, \\
B = \me^{-\mi \theta}
    \begin{pmatrix} \mu_{11} & \mu_{12} \\  \mu_{21} & \mu_{22} 
    \end{pmatrix}.
\end{gather}
\end{subequations}
An element of SU$(2)$ group can be expressed as
\begin{equation}
U = \begin{pmatrix} \alpha & \beta \\  -\bar{\beta} & \bar{\alpha} 
\end{pmatrix},\quad |\alpha|^2 + |\beta|^2 = 1   
\label{SU2}
\end{equation}
and an element of U$(2)$ may include a multiplier 
$\me^{-\mi \theta}$. 

The additional requirement about two SU$(2)$ matrices with the unit determinant  
or $\det(A)=\det(B) = \me^{-\mi \theta}$ for the more general case with U$(2)$
is not related with discussed earlier properties of links between nodes in the chain. 
Such a subtlety is essential further for consideration of effective simulations.

For example, an exchange or the {\it SWAP} gate 
\begin{equation}
\Op{P} = \begin{pmatrix}
1 & 0 & 0 & 0 \\
0 & 0 & 1 & 0 \\
0 & 1 & 0 & 0 \\
0 & 0 & 0 & 1 
\end{pmatrix}
\label{exch} 
\end{equation} 
does not satisfy the requirement about equality of determinants.
The permission to include {\it SWAP} would expand matchgates from nearest
neighbors to arbitrary pairs of qubits, but the extended set of quantum 
gates is universal with lack of possibility for the effective classical 
simulations \cite{Joz8,Joz9}.

An alternative is some ``signed'' {\it SWAP} such as
\begin{equation}
\Op{P}_- = \begin{pmatrix}
1 & 0 & 0 & 0 \\
0 & 0 & 1 & 0 \\
0 & 1 & 0 & 0 \\
0 & 0 & 0 & -1 
\end{pmatrix}
\label{exchm} 
\end{equation}
corresponding to the general case of \Eq{match} with the phase $\theta = \pi$.
Such a matchgate \Eq{exchm} does not change a sign for the states $\ket{\chn k}$
\Eq{qub2scal} with a single unit in the computational basis. 

\bigskip

The model with Hamiltonians clarifies description of some properties \cite{TD2,Joz8}.
For two-dimensional (sub)space of states the exponential expression \Eq{expH} 
generates elements of U$(2)$ for 
decomposition of Hamiltonians with Pauli matrices
\begin{equation}
 \Op H_2(\bm \lambda) = \lambda_0 \OId_2 
  + \lambda_1 \Op{\sigma}^x + \lambda_2 \Op{\sigma}^y + \lambda_3 \Op{\sigma}^z
\label{HU2}
\end{equation}
for real $\lambda_k \in \RR$. Here $\OId_2$ is $2 \times 2$ identity matrix and the
requirement about the unit determinant for SU$(2)$ corresponds to $\lambda_0 = 0$. 

Hamiltonians in \Eq{expH} for generating of matrices, satisfying 
both \Eq{tran4x4} and \Eq{matchtran} can be expressed as
\begin{equation}
\Op H_4(\theta,\bm \lambda) =  \lambda_0 \OId_4
 + \lambda_1 \Op\Sigma + \lambda_2 \Op\Lambda + \lambda_3 \Op\Delta + \theta \Op\Delta',
\label{HM4}
\end{equation} 
where $\OId_4$ is $4 \times 4$ identity matrix and 
the terms in \Eq{HM4} can be represented with Pauli matrices
\begin{subequations}
    \label{HM4sig}
\begin{equation}
\Op\Sigma = \frac{\Op\sigma^x \otimes \Op\sigma^x + \Op\sigma^y \otimes \Op\sigma^y}{2} =
\begin{pmatrix}
0 & 0 & 0 & 0 \\
0 & 0 & 1 & 0 \\
0 & 1 & 0 & 0 \\
0 & 0 & 0 & 0 
\end{pmatrix},
\label{XX+YY} 
\end{equation}
\begin{equation}
\Op\Lambda = \frac{\Op\sigma^y \otimes \Op\sigma^x - \Op\sigma^x \otimes \Op\sigma^y}{2} =
\begin{pmatrix}
0 & 0 & 0 & 0 \\
0 & 0 & -\mi & 0 \\
0 & \mi & 0 & 0 \\
0 & 0 & 0 & 0 
\end{pmatrix}
\label{YX-XY}, 
\end{equation}
\begin{equation}
\Op\Delta = \frac{\Op\sigma^z \otimes \OId_2 - \OId_2 \otimes \Op\sigma^z}{2} =
\begin{pmatrix}
0 & 0 & 0 & 0 \\
0 & 1 & 0 & 0 \\
0 & 0 & -1 & 0 \\
0 & 0 & 0 & 0 
\end{pmatrix}
\label{ZI-IZ}, 
\end{equation}
\begin{equation}
\Op\Delta' = \frac{\Op\sigma^z \otimes \OId_2 + \OId_2 \otimes \Op\sigma^z}{2} =
\begin{pmatrix}
1 & 0 & 0 & 0 \\
0 & 0 & 0 & 0 \\
0 & 0 & 0 & 0 \\
0 & 0 & 0 & -1 
\end{pmatrix}.
\label{ZI+IZ} 
\end{equation}
\end{subequations}
The Hamiltonian for a more general matrix from \Eq{tran4x4} without the requirement
about the unit determinant may also include the term
\begin{equation}
\Op\Theta = \frac{\OId_4 - \Op\sigma^z \otimes \Op\sigma^z}{2} =
\begin{pmatrix}
0 & 0 & 0 & 0 \\
0 & 1 & 0 & 0 \\
0 & 0 & 1 & 0 \\
0 & 0 & 0 & 0 
\end{pmatrix}.
\label{II-ZZ} 
\end{equation}
Finally, the Hamiltonian for the most general matchgate \Eq{match} would require two additional terms
\begin{subequations}
    \label{HMext}
    \begin{equation}
    \Op\Sigma' = \frac{\Op\sigma^x \otimes \Op\sigma^x - \Op\sigma^y \otimes \Op\sigma^y}{2} =
    \begin{pmatrix}
    0 & 0 & 0 & 1 \\
    0 & 0 & 0 & 0 \\
    0 & 0 & 0 & 0 \\
    1 & 0 & 0 & 0 
    \end{pmatrix},
    \label{XX-YY} 
    \end{equation}
    \begin{equation}
    \Op\Lambda' = \frac{\Op\sigma^y \otimes \Op\sigma^x + \Op\sigma^x \otimes \Op\sigma^y}{2} =
    \begin{pmatrix}
    0 & 0 & 0 & -\mi \\
    0 & 0 & 0 & 0 \\
    0 & 0 & 0 & 0 \\
    \mi & 0 & 0 & 0 
    \end{pmatrix}
    \label{YX+XY}, 
    \end{equation}
\end{subequations} 

For a single link with two qubits relations between Hamiltonians and quantum gates 
can be simplified in such representations due to analogies with the single qubit case using correspondence 
between Pauli matrices $(\Op\sigma^x,\Op\sigma^y,\Op\sigma^z)$ and triples 
$(\Op\Sigma,\Op\Lambda,\Op\Delta)$ or $(\Op\Sigma',\Op\Lambda',\Op\Delta')$.

The notation such as \Eq{HM4sig} used for the triples does not look
uniform with respect to all three Pauli matrices, but it is intended to 
indicate some properties of the whole chain. 
Indeed, let us rewrite the {\em spin exchange} 
operator \Eq{exch}
\begin{equation}
\Op{P} =\frac{\OId_4 + \Op\sigma^x \otimes \Op\sigma^x + 
    \Op\sigma^y \otimes \Op\sigma^y + \Op\sigma^z \otimes \Op\sigma^z}{2}.
\tag{\ref{exch}$'$}
\label{exch'} 
\end{equation}
The expression \Eq{exch'} is isotropic, {\em i.e.}, invariant with respect to a 3D rotation $\mt R$
\begin{equation}
\Op\sigma^\alpha \mapsto \sum_{\alpha,\beta} \mt R_{\alpha\beta} \Op\sigma^\beta,\quad \alpha,\beta = x,y,z.
\label{rots3D}
\end{equation}
Four terms \Eq{HM4sig} are anisotropic due to the  preferred axis $z$ in the computational basis, 
but a reduced symmetry still presents
with respect to 2D rotations in $xy$ plane around $z$.

For $\Op\Sigma$ and $\Op\Lambda$ in \Eqs{\ref{XX+YY}, \ref{YX-XY}} such an invariance may be
formally derived from the conservation of 2D length and area respectively. 
Two other terms \Eq{HMext} do not have even such a reduced symmetry. 

A particular case of $\pi/2$ rotation around $z$-axis is essential for the further applications
and denoted here as  
\begin{equation}
\spn J \colon 
(\Op\sigma^x_k, \Op\sigma^y_k,  \Op\sigma^z_k) \mapsto (-\Op\sigma^y_k, \Op\sigma^x_k, \Op\sigma^z_k).
\label{Jrot} 
\end{equation}
Such a transformation does not affect \Eq{HM4sig}, but it changes signs of terms in \Eq{HMext}.

\smallskip

For the representation with Hamiltonians such as \Eq{HM4} an evolution of the whole chain 
is described by the exponent \Eq{expH} with the sum or the linear 
combination of the terms for all existing links.  
The term $\Op\Sigma$ in \Eq{XX+YY} corresponds to the Heisenberg 
XY spin chain and quite common in the models of the perfect state transport
\cite{Perf04,Perf05} and quantum computing \cite{plinko}.
The pure imaginary term \Eq{YX-XY} can be treated as a ``chiral'' \cite{Biam13,Lloyd16}
or ``spiral'' \cite{KT67} and may appear due to the Dzyaloshinskii-Moriya interaction \cite{DVKB06}.
A model with the local magnetic field has additional terms $\Op\sigma_z$ acting on single 
qubits \cite{Perf04,Perf05,plinko} related with \Eqs{\ref{ZI-IZ}, \ref{ZI+IZ}}  
due to the obvious grouping $\Op\Delta\pm\Op\Delta'$.
Finally, the Hamiltonian for a chain with terms \Eq{HM4sig} may be written as 
\begin{eqnarray}
\Op H &=&  \sum_{k=1}^{n-1} \frac{\alpha_k}{2} (\X k \X{k+1} + \Y k \Y{k+1})\nonumber \\
&+& \sum_{k=1}^{n-1} \frac{\beta_k}{2} (\Y k \X{k+1} - \X k \Y{k+1}) 
+ \sum_{k=1}^n \delta_k \Z k.
\label{Ham}
\end{eqnarray}

\smallskip

The alternative model with gates discussed earlier has some subtlety, 
when a result of a step may not be expressed naturally because of 
noncommuting operators. In such a case 
$$\exp(\Op A + \Op B) \neq \exp(\Op A)\exp(\Op B)$$
and a sum of Hamiltonians acting on 
overlapped pairs of qubits after the exponentiation \Eq{expH} in general produces 
$n$-qubit operator without obvious relation with gates for initial pairs. 
Decomposition of such operator on two-qubit gates may be a difficult task. 

A formal resolution of the problem is 
the decomposition on different steps with only commuting operators in each one. 
The partition of a graph on links without common nodes is an easy way 
to ensure commutativity. 
The alternative expression with two operators \Eq{QW} for coined quantum walk 
is a simple example. The approach with partitions is widely used
for discrete time quantum walks \cite{Szeg,stag1,stag}. 

The application of a similar model for qubit chains may be considered
as a special case of quantum cellular automata with the Margolus 
partitioning scheme \cite{SW04}.

\section{Spinoral evolution of qubit chain}
\label{Sec:descr}

\subsection{Clifford algebras and Spin groups}
\label{Sec:spin}

The matchgates \Eq{match} were already adapted above \Eq{matchtran} together
with the matrix \Eq{tran4x4} used for the description of transition
between scalar and qubit chains. The nearest neighbor matchgates 
were exploited for the description of quantum circuits effectively modeled 
on a classical computer \cite{Val1,TD2,Kni1,Joz8,Joz9,JMS15,Brod16}.
The term ``matchgates'' is used here for compatibility 
with other works and some historical reasons, but further 
methods rely on rather standard theory of Clifford algebras and 
Spin groups \cite{SLang,ClDir}.

The Hamiltonians for nearest neighbor matchgates can be expressed 
with $2n$ anticommuting generators of the Clifford algebra \cite{TD2,Joz8} 
also known due to the Jordan-Wigner transformation \cite{JW}. The relation 
corresponds to the standard representation of Spin groups \cite{Vla0,ClDir}.

The similar approach also appears earlier due to natural analysis of 
universality using Hamiltonians of quantum gates \cite{DiVin95}, because
a non-universal set of gates was directly related with Clifford 
algebras and Spin groups of multidimensional Euclidean spaces  
\cite{Kit00,BK00,Vla0,Wilc09}.

The Clifford algebra $\Cl(m)$ of $m$-dimensional Euclidean space is defined by
$m$ generators $\e_k$ with relations \cite{ClDir}
\begin{equation}
 \e_j\e_k + \e_k\e_j = -2 \delta_{jk}\Id,
 \quad k,j=1,\ldots,m 
\label{ClDef}
\end{equation}
The $2^m$-dimensional algebra $\Cl(m)$ is spanned by different products of $\e_k$. 
The linear span of generators $\e_k$ maps initial Euclidean space 
into $m$-dimensional subspace $\mathcal V$ of $\Cl(m)$ and due to \Eq{ClDef} 
the Euclidean norm of $\bm v \in \mathcal V$ satisfies $|\bm v|^2 = -\bm v^2$.

The Spin$(m)$ group is defined by all possible products with {\it even} number of 
elements from $\mathcal V$ with unit norm. The products
of {\it arbitrary} number of such elements define group Pin$(m)$ \cite{ClDir}.

The important property of Spin group is relation with group of rotations
SO$(m)$, because for any $\mt R \in \mathrm{SO}(m)$ the rotation 
\begin{equation}
\bm v' = \mt R \bm v, \quad
\bm v'_k = \sum_{j=1}^m \mt R_{kj}\bm v_j
\label{rotsD}
\end{equation}
may be rewritten as the {\em adjoint action}
\begin{equation}
\bm v' = \spn S_{\mt R} \bm v \spn S^{-1}_{\mt R} 
\equiv \Ad_{S_{\mt R}}(\bm v), 
\quad \spn S_{\mt R} \in \mathrm{Spin}(m).
\label{rotspin}
\end{equation}
Here two elements $\pm\spn S_{\mt R}$ correspond to the same rotation $\mt R$
and \Eq{rotspin} defines 2-fold homomorphism, 
{\em i.e.,} a map respecting the composition of transformations.
For effective simulations further is essential a reciprocal
opportunity to use rotations for the work with quantum circuits 
represented by the Spin group. 
For the generators from \Eq{rotsD} and \Eq{rotspin} follows
\begin{equation}
\spn S_{\mt R} \e_k \spn S^{-1}_{\mt R} = \sum_{j=1}^m \mt R_{kj}\e_j.
\label{rotgen}
\end{equation}

In even dimension $m=2n$ the generators can be expressed with the Jordan-Wigner
method \cite{JW,ClDir} 
\begin{subequations}
    \label{ekn}
    \begin{eqnarray}
    \e_{k} & = &
    \mi\,{\underbrace{\Op\sigma^z\otimes\cdots\otimes \Op\sigma^z}_{k-1}\,}\otimes
    \Op\sigma^x\otimes\underbrace{\OId_2\otimes\cdots\otimes\OId_2}_{n-k} \, ,
    \label{ek1}\\
    \e_{k+n} & = &
    \mi\,{\underbrace{\Op\sigma^z\otimes\cdots\otimes \Op\sigma^z}_{k-1}\,}\otimes
    \Op\sigma^y\otimes\underbrace{\OId_2\otimes\cdots\otimes\OId_2}_{n-k} \, ,
    \label{ek2}
    \end{eqnarray}
\end{subequations}
where $k = 1,\ldots,n$.
An alternative short notation is useful further
\begin{equation}\label{eXYZ}
 \e_{k} = \mi\, \Z 1 \cdots \Z{k-1} \X k,\quad
\e_{k+n}  = \mi\, \Z 1 \cdots \Z{k-1} \Y k .
\end{equation}

The expressions \Eq{HM4sig} for two neighboring qubits may be rewritten
with \Eq{eXYZ}
\begin{subequations}
    \label{ClifPair}
\begin{align}\label{XX+YY_k} 
\Op\Sigma_{k,k+1} 
&=\frac{\e_{k}\e_{k+n+1}+\e_{k+1}\e_{k+n}}{2\mi}, \\ 
\label{YX-XY_k}
\Op\Lambda_{k,k+1}
&=\frac{\e_{k}\e_{k+1}+\e_{k+n}\e_{k+n+1}}{2\mi}, \\
\label{ZI-IZ_k}
\Op\Delta_{k,k+1} 
&=\frac{\e_{k}\e_{k+n}-\e_{k+1}\e_{k+n+1}}{2\mi}, \\
\label{ZI+IZ_k} 
\Op\Delta'_{k,k+1} 
&=\frac{\e_{k}\e_{k+n}+\e_{k+1}\e_{k+n+1}}{2\mi}. 
\end{align}
\end{subequations}

Supplementary Hamiltonians \Eq{HMext} needed for the representation
of arbitrary matchgates can be rewritten as
\begin{subequations}
    \label{ClifExt}
    \begin{align}\label{XX-YY_k} 
    \Op\Sigma'_{k,k+1} 
    & =\frac{\e_{k}\e_{k+n+1}-\e_{k+1}\e_{k+n}}{2\mi},\\
    \label{YX+XY_k} 
    \Op\Lambda'_{k,k+1}
    & =\frac{\e_{k}\e_{k+1}-\e_{k+n}\e_{k+n+1}}{2\mi},
    \end{align}
\end{subequations}

An analogue of $\Op\Theta$ \Eq{II-ZZ} may not be expressed in a similar way,
because it requires four Clifford generators
\begin{equation}
\Op\Theta_{k,k+1} 
= \frac{\Id  + \e_{k}\e_{k+n}\e_{k+1}\e_{k+n+1}}{2}.
\label{I-ZZ_k} 
\end{equation}

The exchange operator $\Op{P}$ \Eq{exch'} also requires four Clifford generators,
but a ``signed'' version $\Op{P}_-$ might be expressed using \Eq{ClifPair}. 

\smallskip

It may be convenient to consider a more general definition for \Eq{ClifPair} 
\begin{subequations}
    \label{ClifPair2}
    \begin{align}\label{XX+YY_kj} 
    \Op\Sigma_{k,j} 
    &=\frac{\e_{k}\e_{j+n}+\e_{j}\e_{k+n}}{2\mi}, \\ 
    \label{YX-XY_kj}
    \Op\Lambda_{k,j}
    &=\frac{\e_{k}\e_{j}+\e_{k+n}\e_{j+n}}{2\mi}.
    \end{align}
\end{subequations}
Both $\Op\Delta_{k,k+1}$ 
and $\Op\Delta'_{k,k+1}$ may be expressed in terms of \Eq{XX+YY_kj}
due to the identity
\begin{equation}
\Z{k}=-\mi\,\e_{k}\e_{k+n} = \Op\Sigma_{k,k}.
\label{eeZk}
\end{equation}

\subsection{Admissible evolution of qubit chain}
\label{Sec:admi}

The evolution of a qubit chain considered above preserves the subspace spanned by states 
$\ket{\chn k}$ \Eq{qub2scal} with a single unit in the computational basis.
Linear combinations of Hamiltonians \Eq{HM4sig} for nearest neighbor qubits
generate transformations respecting such a subspace and compositions 
of corresponding quantum gates \Eq{tran4x4} also have the necessary property. 
For the certainty the term {\em admissible} is used further for such an evolution, 
relevant quantum gates, Hamiltonians and elements of Spin group. 

Let us use expressions \Eq{ClifPair} with elements of the Clifford algebra.
A replacement of $n$ pairs
\begin{equation} 
\label{Jsub}
 \e_{k} \mapsto  -\e_{k+n}, \quad \e_{k+n} \mapsto \e_{k}, \quad k = 1,\ldots,n
\end{equation}
does not change operators \Eq{ClifPair}, but it alternates signs in \Eq{ClifExt}.
Such properties may be simply checked using the analogue transformation
with Pauli matrices introduced earlier \Eq{Jrot}.

The substitution \Eq{Jsub} can be treated as $2n \times 2n$ matrix
\begin{equation}
\mt J = \begin{pmatrix}
\mt 0_n & \mt 1_n & \\
-\mt 1_n & \mt 0_n
\end{pmatrix},
\label{MatJ} 
\end{equation}
where $\mt 0_n$, $\mt 1_n$ denote $n \times n$ zero and unit matrices, respectively.
Because $\mt J$ is orthogonal matrix, \Eq{Jsub} also may be rewritten as \Eq{rotspin}
for the adjoint action $\Ad_{\spn J}$  with the element $\spn J$ of the Spin group 
expressed as the composition of $n$ elementary terms derived from \Eq{Jsub}
\begin{equation}
    \spn J =  \frac{1}{\sqrt{2^n}}\prod_{k=1}^n(\Id - \e_{k}\e_{k+n}) =
    \frac{1}{\sqrt{2^n}}\prod_{k=1}^n(\OId + \mi\,\Z k).
    \label{SJ}
\end{equation}

Admissible Hamiltonians $\Op H_{\rm a}$ for the evolution of a chain
are expressed as linear combinations 
of elements \Eq{ClifPair}. Such Hamiltonians satisfy 
$\Op H_{\rm a} = \Ad_{\spn J}(\Op H_{\rm a}) = \spn J \Op H_{\rm a} \spn J^{-1}$, 
{\em i.e.}, they commute with the element $\spn J$
\begin{equation}\label{Hchain}
\Op H_{\rm a} \spn J = \spn J \Op H_{\rm a}.
\end{equation}
Due to \Eq{expH} the evolution of a quantum chain generated by such Hamiltonians also 
commutes with $\spn J$ and the same is true for the particular case with nearest neighbor 
quantum gates such as \Eq{tran4x4} and for any circuit composed from them.
Operators $\Op U_{\rm a}$ describing admissible evolution of a qubit chain
due to such quantum gates and circuits also commute with $\spn J$
\begin{equation}\label{Uchain}
\Op U_{\rm a} \spn J = \spn J \Op U_{\rm a}. 
\end{equation}

Some properties of $\spn J$ may be more convenient to explain using an operator
\begin{equation}
\label{Nz}
\Op N^z = \sum_{k=1}^n \frac{\OId-\Z k}{2} = \frac{n}{2}\OId-\frac{1}{2}\sum_{k=1}^n \Z k.
\end{equation}
Any vector of the computational basis $\ket{\Psi}$ meets 
\begin{equation}
\Op N^z \ket{\Psi} = N_\Psi\ket{\Psi},
\label{NzEVec}
\end{equation}
where $N_\Psi$ is the number of units in a binary notation.

The relation between $\Op N^z$ and $\spn J$ may be 
expressed using \Eq{SJ} and \Eq{Nz} 
\begin{equation}
\me^{\mi\frac{\pi}{4}n}\me^{-\mi\frac{\pi}{2}\Op N^z}
= \prod_{k=1}^n\me^{\mi\frac \pi 4 \Z k} 
= \prod_{k=1}^n\frac{\OId + \mi\,\Z k}{\sqrt 2} 
= \spn J
\label{SJNz}
\end{equation}

Due to \Eq{NzEVec} $\Op N^z \ket{\chn k} = \ket{\chn k}$ 
for the basis of $n$-dimensional subspace \Eq{qub2scal}  and the same
is true for any elements of the subspace $\ket{\chn \psi}$ represented as a linear 
combination of the basic states
\begin{equation}
 \ket{\chn \psi} = \sum_{k=1}^n \psi_k \ket{\chn k}, \quad
 \Op N^z \ket{\chn\psi} = \ket{\chn\psi}
\label{chnpsi}
\end{equation}

The similar expressions \Eq{NzEVec} are valid for any subspace spanned 
by basic vectors with the fixed number of units in the computational basis.
An operator $\Op U_{\rm a}$ respecting the numbers acts irreducibly on all such 
subspaces and commutes with both $\Op N^z$ and $\spn J$ due to \Eq{SJNz}. 

\subsection{Annihilation and creation operators}
\label{Sec:CAR}

The admissible transformations described above correspond to some
subgroup of the Spin group related with $2n \times 2n$ orthogonal 
matrices of rotations discussed earlier \Eq{rotspin}. 
The matrices are also {\em symplectic} due to commutation with $\mt J$ \Eq{MatJ} 
and they are belong to {\em symplectic orthogonal group} OSp$(2n)$ \cite{Post}.

A crucial property of OSp$(2n)$ is the isomorphism with the special unitary group SU$(n)$
\cite{Post}. 
Let us consider a correspondence between complex and real matrices 
written
\begin{equation}
 \mt M_\CC = \mt M_\Re + \mi \mt M_\Im \longleftrightarrow 
 \mt M_\RR =\begin{pmatrix}
  \mt M_\Re & \mt M_\Im & \\
 -\mt M_\Im & \mt M_\Re
 \end{pmatrix}
\label{MatR2C}
\end{equation}
where $\mt M_\Re$ and $\mi\mt M_\Im$ are $n \times n$ matrices composed respectively from real and 
imaginary parts of elements $\mt M_\CC$. 

For the unitary matrix
$\mt U = \mt U_\Re + \mi \mt U_\Im \in \mathrm{SU}(n)$ the correspondence
\Eq{MatR2C} produces the standard isomorphism \cite{Post} with 
$\mt M_\RR = \mt R_{\mt U} \in \mathrm{OSp}(2n)$ 

\begin{equation}\label{UOSp}
 \mt R_{\mt U}  =  \begin{pmatrix} 
 \mt U_\Re & \mt U_\Im & \\  -\mt U_\Im & \mt U_\Re 
 \end{pmatrix}. 
\end{equation} 

For the exponential representation $\mt U = \exp(\mi \mt H \tau)$ with 
a Hermitian matrix $\mt H$ the same procedure \Eq{MatR2C} should be applied to  
$\mt M_\CC =\mi\mt H$ producing a generator for an appropriate rotation.

The approach is also related with the
Jordan-Wigner operators \cite{Weyl,JW,ClDir}  
\begin{equation}
\Op a_k = \frac{\e_k+\mi\e_{k+n}}{2\mi},\quad
\Op a_k^\dag = \frac{\e_k-\mi\e_{k+n}}{2\mi}.
\label{creann}
\end{equation}

The \Eq{creann} meet the {\em canonical anticommutation relations} for fermionic annihilation
and creation operators 
\begin{equation}
\{\Op a_j,\Op a_k^\dag\} = \delta_{jk},\quad
\{\Op a_j,\Op a_k\} = \{\Op a_j^\dag,\Op a_k^\dag\}=0
\label{CAR}
\end{equation}

The approach to the effective simulation of quantum circuits based on these
operators may be found in \cite{TD2}, but few
points should be clarified here.

Transition from generators $\e_k$ to operators \Eq{creann} can be expressed
formally by $2n \times 2n$ complex matrix 
\begin{equation}
    \mt\Xi =
    \frac{1}{2}\begin{pmatrix} 
        \mt 1_n & \mi \mt 1_n & \\  \mt 1_n & -\mi\mt 1_n 
    \end{pmatrix}.
    \label{MatXi} 
\end{equation}
For $\mt R \in \mathrm{OSp}(2n)$ the matrix
$\mt U_\Xi$ produced by transformation \Eq{MatXi} can be expressed 
directly using the unitary matrix $\mt U$ introduced earlier \Eq{UOSp}
\begin{equation}
\mt U_\Xi = \mt{\Xi R_U \Xi}^{-1} 
=  \begin{pmatrix} 
\mt U_\Re + \mi\mt U_\Im & \mt 0_n\\ \mt 0_n & \mt U_\Re -\mi\mt U_\Im
\end{pmatrix} 
=  \begin{pmatrix} 
 \mt U & \mt 0_n\\ \mt 0_n & \bar{\mt U}
\end{pmatrix}
\label{dblU}
\end{equation} 

In such representation the requirements about commutativity with $\mt J$ 
become rather trivial due to diagonalization of the matrix
\begin{equation}
\mt J_\Xi = \mt{\Xi J \Xi}^{-1} =
\mi\begin{pmatrix}
\mt 1_n & \mt 0_n & \\
\mt 0_n & -\mt 1_n
\end{pmatrix}
\label{MatIXi} 
\end{equation}
obviously commuting with any matrix \Eq{dblU}.

Let us consider for the admissible chain evolution $\Op{\spn U} \equiv \spn S_{\mt R}$
transformations of operators \Eq{creann}
\begin{equation}
\Op a'_k = \Op{\spn U} \Op a_k \Op{\spn U}^{-1},\quad
\Op a'{}^\dag_k = \Op{\spn U} \Op a^\dag_k \Op{\spn U}^{-1}.
\label{Uspin}
\end{equation}
Due to \Eq{rotsD} and \Eq{rotgen} it corresponds 
to formal complex transformations 
$\bm a = \mt\Xi \bm v$, $\bm a' = \mt\Xi \bm v'$ and
\begin{equation}
\bm a' = \mt{\Xi R \Xi}^{-1} \bm a = \mt U_\Xi \bm a.
\label{aUXi}
\end{equation}
Finally, the transformation \Eq{Uspin} for the admissible evolution of the chain can be 
expressed as an analogue of \Eq{rotgen}
\begin{equation}
\Op{\spn U} \Op a_k \Op{\spn U}^\dag = \sum_{j=1}^n \mt U_{kj}\Op a_j, \quad
\Op{\spn U} \Op a_k^\dag \Op{\spn U}^\dag = \sum_{j=1}^n \bar{\mt U}_{kj}\Op a_j^\dag .
\label{rotsU}
\end{equation}
similar with the {\em restricted case} of classical simulations considered
in \cite{TD2}.

\smallskip

There is some analogy with linear optics also described by OSp$(2n)$ 
transformations of bosonic creation and annihilation operators discussed 
elsewhere \cite{Brod16,pbl}.

\smallskip

Quadratic expressions for Hamiltonians \Eq{ClifPair2} also may be rewritten 
using fermionic operators  
\begin{subequations}
    \label{FermPair}
    \begin{align}\label{FermPairX} 
        \Op\Sigma_{k,j} 
        &=\frac{\Op a_{k}\Op a_{j}^\dag+\Op a_{j}\Op a_{k}^\dag}{2}, \\ 
        \label{FermPairY}
        \Op\Lambda_{k,j}
        &=\frac{\Op a_{k}\Op a_{j}^\dag-\Op a_{j}\Op a_{k}^\dag}{2\mi}.
    \end{align}
\end{subequations}

\section{Effective modeling of qubit chains}
\label{Sec:effmod}

\subsection{Single-particle simulation}
\label{Sec:sing}

The effective classical simulation of quantum circuits discussed here may use
an expression \Eq{rotgen} with earlier developed methods \cite{Joz8,Joz9} 
almost without modifications and decompositions \Eq{rotsU} also
provides an alternative approach similar with discussed in \cite{TD2}.

Let us consider a scheme appropriate for many cases discussed
earlier \cite{TD2,Joz8,Joz9,Brod16}. A quantum circuit is described 
by the unitary operator $\Op{\spn U}$ and composed from products of
separate gates. It is applied to the initial state $\ket{\psi_\mathrm{in}}$ with
the final measurement of probabilities defined by some operator $\Op M_\mathrm{out}$
\begin{equation}
p = \bra{\psi_\mathrm{in}} \Op{\spn U}^\dag \Op M_\mathrm{out} \Op{\spn U} \ket{\psi_\mathrm{in}} 
= \Tr(\Op M_\mathrm{out} \Op{\spn U} \Op\rho_\mathrm{in} \Op{\spn U}^\dag),
\label{PQgen}
\end{equation}
where $\Op\rho_\mathrm{in} = \ket{\psi_\mathrm{in}}\bra{\psi_\mathrm{in}}$ 
is the density operator of the initial state. The expression with trace $\Tr$ can be
also used for a mixed initial state. 
Sometimes the probability is characterized
by projector to ``out state'' $\ket{\psi_\mathrm{out}}$ \cite{TD2,Brod16} and in such a
case $\Op M_\mathrm{out} = \ket{\psi_\mathrm{out}}\bra{\psi_\mathrm{out}}$
may be used in \Eq{PQgen} producing an equivalent expression 
$p = |\bra{\psi_\mathrm{out}} \Op{\spn U} \ket{\psi_\mathrm{in}}|^2,$ 
but more general $\Op M_\mathrm{out}$ is considered here to take into account
an alternative approach \cite{Joz8,Joz9} and methods discussed below.

\medskip 

Let us consider for modeling of a qubit chain initial states $\ket{\chn j}$ \Eq{qub2scal}  
and measurement operators
\begin{equation}
\chn{\Op{M}\!}_k  \equiv  \Op n^z_k =
\frac{\OId-\Op\sigma^z_k}{2}  = \Op a_k^\dag\Op a_k.
\label{Mkaa}
\end{equation}

With the operator $\chn{\Op{M}\!}_k$ the probability to find the unit at the node $k$ for the initial state
$\ket{\chn{j}}$ may be found using \Eq{PQgen} and \Eq{Mkaa} 
\begin{eqnarray}
p_{j\to k} &=& \bra{\chn{j}} \Op{\spn U}^\dag \Op a_k^\dag\Op a_k \Op{\spn U} \ket{\chn{j}} 
 = \bra{\chn{j}} \Op{\spn U}^\dag \Op a_k^\dag \Op{\spn U}  \, \Op{\spn U}^\dag \Op a_k \Op{\spn U} \ket{\chn{j}}\nonumber\\
  &=&\sum_{l,r} \bra{\chn{j}}\bar{\mt U}_{kl}^*\Op a_l^\dag\mt U_{kr}^* \Op a_r\ket{\chn{j}}
  =\sum_{l,r} \mt U_{lk}\bar{\mt U}_{rk} \bra{\chn{j}} \Op a_l^\dag\Op a_r\ket{\chn{j}} \nonumber\\
  &=&\sum_l |\mt U_{lk}|^2 \bra{\chn{j}} \Op a_l^\dag\Op a_l\ket{\chn{j}} 
  = |\mt U_{jk}|^2.
\label{pMjk}
\end{eqnarray}

The evolution of states $\ket{\chn j}$ can be described more directly due to 
yet another approach also used in \cite{TD2}. Let us denote
\begin{equation}
\ket{\chn\vac} \equiv \ket{\underbrace{0\ldots 0}_n}.
\label{zeros}
\end{equation}
Creation operators $\Op a_k^\dag$ defined in \Eq{creann} 
for the representation \Eq{eXYZ} meet the natural condition
\begin{equation}
\ket{\chn k} = \Op a_k^\dag \ket{\chn\vac}.
\label{cre}
\end{equation}
Any admissible evolution conserves the number of units in the computation basis and so 
\begin{equation}\label{U0}
\Op{\spn U}\ket{\chn\vac} = \ket{\chn\vac}.
\end{equation}
Using such properties it may be written
\begin{eqnarray}
\Op{\spn U} \ket{\chn{k}} &=& \Op{\spn U} \Op a_k^\dag \ket{\chn\vac}
= \Op{\spn U} \Op a_k^\dag \Op{\spn U}^\dag \ket{\chn\vac} \nonumber\\
&=& \sum_{j=1}^n \bar{\mt U}_{kj}\Op a_j^\dag \ket{\chn\vac}
= \sum_{j=1}^n \bar{\mt U}_{kj} \ket{\chn j}
= \sum_{j=1}^n \mt U^\dag_{jk} \ket{\chn j}.
\label{evolk}
\end{eqnarray}
For a state $\ket{\chn\psi}$ defined as a linear superposition of $\ket{\chn{k}}$  
\begin{equation}
 \Op{\spn U} \ket{\chn\psi}
  =  \sum_{k=1}^n\psi_k \Op{\spn U}\ket{\chn{k}}
  = \sum_{j,k=1}^n \mt U^\dag_{jk} \psi_k \ket{\chn{j}}
  \equiv \sum_{j=1}^n \psi'_j \ket{\chn{j}},\quad
  \psi'_j = \sum_{k=1}^n \mt U^\dag_{jk} \psi_k.
\label{evolpsi}  
\end{equation}

Formally, the \Eq{evolpsi} produces a correspondence with the unitary 
evolution ${\mt U}^\dag$ of a scalar chain with $n$ nodes considered earlier.
It has some similarity with the approach used for the perfect state transfer
for particular Hamiltonians \Eq{HSx} or \Eq{HSxyz} of a higher-spin system. 
However, ${\mt U}$ in \Eq{rotsU} is an arbitrary $n\times n$ unitary matrix
and the evolution of state $\ket{\chn\psi}$ due to \Eq{evolpsi} can be
considered for many different kinds of quantum chains with required properties. 

The simulation of the measurement for a qubit chain in single-particle is effective, 
because the evolution is limited by $n$-dimensional span of states $\ket{\chn k}$ 
\Eq{qub2scal} and a scalar chain with $n$ nodes can be used instead as a model
without lost of generality. Such a correspondence also clarifies \Eq{pMjk} 
derived earlier less directly.

\subsection{Multi-particle simulation}
\label{Sec:mult}

Rather straightforward transition to the consideration of an evolution with many particles 
is an essential property of the qubit chain model considered here.
The general scheme from \cite{TD2,Joz8,Joz9,Brod16} is again appropriate 
for such a purpose. 

The generalization of a single-particle configuration \Eq{cre} is a basic state  
\begin{equation}
\ket{\chn K} \equiv
\ket{\chn {k_1,\ldots,k_m}} = \Op a_{k_1}^\dag\!\cdots \Op a_{k_m}^\dag \ket{\chn\vac}
\label{crem}
\end{equation}
with $m$ units in positions $k_1 < \cdots < k_m$. For example,
an analogue of \Eq{qub2scal} with {\em two neighboring} particles is
\begin{equation}
\ket{\chn{k,k+1}} \equiv \ket{{\underbrace{0\ldots 0}_{k-1}\,}11\underbrace{0\ldots 0}_{n-k-1}}
=\Op a_{k}^\dag\Op a_{k+1}^\dag \ket{\chn\vac}.
\label{cre2}
\end{equation}

Let us denote
\begin{equation}\label{au}
\Op a^u_k = \Op{\spn U} \Op a_k \Op{\spn U}^\dag = \sum_l \mt U_{kl} \Op a_l  
\end{equation} 
with the obvious property $\Op a^u_k\Op a^u_l = - \Op a^u_l\Op a^u_k$.
For the initial states \Eq{crem} taking into account \Eq{U0}
\begin{equation}
\Op{\spn U}\ket{\chn K} =
\Op{\spn U}\ket{\chn {k_1,\ldots,k_m}}
= \Op{\spn U}\Op a_{k_1}^\dag\Op{\spn U}^\dag\!\cdots \Op{\spn U}\Op a_{k_m}^\dag\Op{\spn U}^\dag\,
  \Op{\spn U}\ket{\chn\vac}  
= \Op a_{k_1}^{u\dag}\!\cdots \Op a_{k_m}^{u\dag} \ket{\chn\vac}.
\label{Ucrem}
\end{equation}
The evolution is represented as the {\em antisymmetric product} of operators $\Op a_k^{u\dag}$ 
generating distributions \Eq{evolk} for particles in different initial positions.
Such representation should be considered rather as some model {\em without interaction}.

Indeed, let us associate with any single-particle state $\ket{\chn \psi}$ \Eq{chnpsi}
an operator
\begin{equation}\label{opsi}
\Op{\psi} \equiv \sum_{k=1}^n \psi_k \Op a_k^\dag, \qquad
\Op{\psi} \ket{\chn\vac} = \ket{\chn \psi}.
\end{equation}
For the two-particle case the composition of such operators 
correspond to an anti-symmetric (exterior) product defined on
$n$-dimensional single-particle space
\begin{equation}\label{extop}
\Op{\psi} \Op{\phi} \ket{\chn\vac} = \ket{\chn \psi} \wedge \ket{\chn \phi}. 
\end{equation}
It follows directly from the antisymmetry of creation operators and 
the formal definition of the exterior product \cite{SLang,ClDir} for the basic states
\begin{equation}\label{extbas}
  \ket{\chn{j,k}} \equiv 
  \ket{\chn j} \wedge \ket{\chn k} = - \ket{\chn k} \wedge \ket{\chn j}
  \quad (j < k), \qquad \ket{\chn j} \wedge \ket{\chn j} = 0.
\end{equation}
The generalization to the multi-particle space is straightforward and the evolution
\Eq{Ucrem} may be rewritten now as an exterior product of single-particle terms
\begin{equation}\label{extevol}
\Op{\spn U}\ket{\chn K} =
\Op{\spn U}\ket{\chn {k_1,\ldots,k_m}} = 
\Op{\spn U}\ket{\chn k_1} \wedge \cdots \wedge \Op{\spn U}\ket{\chn k_m}.
\end{equation} 
It was already shown above, that each single-particle term in \Eq{extevol}
evolution can be modeled by an operator $\mt U^\dag$ on a simple chain with 
$n$ nodes.
Thus, the multi-particle evolution corresponds to the anti-symmetric product
of $m$ such chains. 

\smallskip

An effective simulation of the evolution together with the measurement may require additional efforts 
if the number of particles is large. Some general methods developed for the description of
match-circuits may be found elsewhere \cite{TD2,Kni1,Joz8,Joz9,JMS15,Brod16} and 
the {\em restricted case} is relevant here for the {\em admissible evolution} with creation 
and annihilation operators and it was also already discussed earlier by different
authors \cite{TD2,Brod16}.

Methods of simulation depend on the scheme of the initialization and the measurement.
For the many problem of the quantum state distribution an initial state  
may be chosen from the computational basis. The output of the simulation
may use an approach from \cite{Joz9} with the measurement in the computational basis. 
In such a case the probability of the ``occupation'' (unit) for any node $k$ may be 
effectively calculated using a simplified approach discussed in \cite{Joz8,Joz9}.

Let us write an analogue of \Eq{pMjk} for the probability to find
a particle on the node $k$ after the evolution of the multi-particle state \Eq{crem} 
\begin{eqnarray}
p_{K\to k} &=& \bra{\chn{K}} \Op{\spn U}^\dag \Op a_k^\dag\Op a_k \Op{\spn U} \ket{\chn{K}} 
=\sum_{l,r} \mt U_{lk}\bar{\mt U}_{rk} \bra{\chn{K}} \Op a_l^\dag\Op a_r\ket{\chn{K}} \nonumber\\
&=&\sum_l |\mt U_{lk}|^2 \bra{\chn{K}} \Op a_l^\dag\Op a_l\ket{\chn{K}} 
=\sum_{l \in K}|\mt U_{lk}|^2.
\label{pMKk}
\end{eqnarray}

In the more general case such approach with the separate measurements 
of a qubits could be not enough to uncover some nontrivial quantum 
correlations between qubits.
Measurements of multi-qubit output would require more complicated
methods for effective classical simulations \cite{TD2,Brod16}.
However, the multi-particle case for 
given model has the understanding exterior structure \Eq{extevol}
and more general measurement schemes are not discussed here.

\section{Conclusion}
\label{Sec:concl}

The application of a particular non-universal set of quantum gates and Hamiltonians
was discussed in this work. Different examples of the state distribution along a chain 
of qubits was investigated for such a purpose.

A single particle on a chain is a convenient simplified model and it is
used for the comparison of scalar and qubit chains in \Sec{cmp}.
An arbitrary unitary operator on a scalar chain
can be associated with some effectively modeled evolution of a qubit chain 
using methods from \Sec{CAR}. 
For the multi-particle case an evolution of a qubit chain for the considered model
is mapped in \Sec{mult} into the anti-symmetric product of such a scalar chains.

The certain difficulty of the considered approach is a lack of the simple possibility for
a generalization of the methods for effective modeling from a chain on arbitrary
graph, because linked nodes may not always correspond to consequent
indexes. 
For a qubit ring the similar approach still may work efficiently
\cite{Brod16}, but the discussion about more general graphs falls outside 
the limits of presented work.

\section*{Acknowledgements}
 The author is grateful to anonymous referees for corrections and suggestions to initial
 version of the article.

\end{document}